\documentclass[aps,prd,twocolumn,nofootinbib,superscriptaddress, nolongbibliography]{revtex4-2}

\usepackage[dvipsnames, usenames]{xcolor}
\definecolor{linkcolor}{rgb}{0.0,0.3,0.5}
\definecolor{dodgerblue}{HTML}{1E90FF}
\usepackage[unicode, colorlinks=true, linkcolor=linkcolor, citecolor=linkcolor, filecolor=linkcolor,urlcolor=linkcolor, pdfusetitle]{hyperref}

\usepackage[utf8]{inputenc}
\usepackage[T1]{fontenc}
\usepackage{amsmath,amssymb,amsfonts}
\usepackage{graphicx}
\usepackage{hyperref}
\usepackage{color}
\usepackage{orcidlink}
\usepackage{physics}
\usepackage{comment}

\usepackage[normalem]{ulem} %

\newcommand{\Tsft}{T_\mathrm{SFT}}
\newcommand{\sft}[1]{{#1}_{\alpha}}

\newcommand{\milan}{\affiliation{Dipartimento di Fisica ``G. Occhialini'', Universit\'a degli Studi di Milano-Bicocca, Piazza della Scienza 3, 20126 Milano, Italy}}
\newcommand{\infn}{\affiliation{INFN, Sezione di Milano-Bicocca, Piazza della Scienza 3, 20126 Milano, Italy}}

\begin{document}

\title{
Ab uno disce omnes: Single-harmonic search for extreme mass-ratio inspirals
}

\author{Lorenzo Speri\orcidlink{0000-0002-5442-7267}}\email{lorenzo.speri@esa.int}
\affiliation{European Space Agency (ESA), European Space Research and Technology Centre (ESTEC), Keplerlaan 1, 2201 AZ Noordwijk, the Netherlands}
\author{Rodrigo Tenorio$\,$\orcidlink{0000-0002-3582-2587}}
\milan \infn
\author{Christian Chapman-Bird$\,$\orcidlink{0000-0002-2728-9612}}
\affiliation{Institute for Gravitational Wave Astronomy \& School of Physics and Astronomy, University of Birmingham, Edgbaston, Birmingham B15 2TT, UK}

\author{Davide Gerosa$\,$\orcidlink{0000-0002-0933-3579}}
\milan \infn

\date{\today}

\begin{abstract}
Extreme mass-ratio inspirals (EMRIs) are one of the key sources of gravitational waves for space-based detectors such as LISA. However, their detection remains a major data analysis challenge due to the signals' complexity and length. 
We present a semi-coherent, time-frequency search strategy for detecting EMRI harmonics without relying on full waveform templates. 
We perform an injection and search campaign of single mildly-eccentric equatorial EMRIs in stationary Gaussian noise.
The detection statistic is constructed solely from the EMRI frequency evolution, which is modeled phenomenologically using a Singular Value Decomposition basis. The pipeline and the detection statistic are implemented in time-frequency, enabling efficient searches over one year of data in approximately one hour on a single Graphic Processing Unit (GPU).
The search pipeline achieves 94\% detection probability at $\mathrm{SNR} = 30$ for a false-alarm probability of $10^{-2}$, recovering the frequency evolution of the dominant harmonic within 1\% relative error. 
By mapping the EMRI parameters consistent with the recovered frequency evolution,
we show that the semi-coherent detection statistic enables a sub-percent precision estimation of the EMRI intrinsic parameters.
These results establish a computationally efficient framework for constructing EMRI proposals for the LISA global fit.

\noindent\emph{Code repository:} \href{https://github.com/lorenzsp/EMRI-Search/}{EMRI-Search}
\end{abstract}

\maketitle

\section{Introduction}
Extreme mass-ratio inspirals (EMRIs) involve the capture and inspiral of a compact stellar remnant into a massive black hole (MBH). These systems are one of the key gravitational-wave (GW) sources of future space-based detectors like the Laser Interferometer Space Antenna (LISA)~\cite{LISA:2024hlh}, TianQin~\cite{Luo:2015ght}, and Taiji~\cite{Hu:2017mde}.
The scientific potential of EMRI observations is vast.
They offer a unique perspective of the properties and environments of MBHs in galactic cores~\cite{Speri:2022upm,Yunes:2011ws,Kocsis:2011dr,Duque:2024mfw,Barausse:2014tra,Kejriwal:2023cqv,Khalvati:2024tzz,Singh:2025pzh,Gair:2008bx,Gair:2004si}.
They also provide an effective means of constraining fundamental physics, being both highly sensitive to effects predicted by modified gravity theories~\cite{Zi:2025lio,Zou:2025fsg,Speri:2024qak,LISA:2022kgy,Piovano:2020ooe,Barausse:2020rsu,Berry:2019wgg,Barack:2018yly,Chua:2018yng,Babak:2017tow,Chamberlain:2017fjz,Berti:2015itd,Gair:2013rwa,Vigeland:2010sr,Sopuerta:2009iy,Barack:2006pq,Ryan:1997hg} and effective probes of cosmological expansion ~\cite{Liu:2023onj,Toscani:2023gdf,Laghi:2021pqk}.
Despite this strong case for EMRI science, identifying their signals in noisy data remains a largely unsolved data analysis problem due to several factors:
\begin{itemize}
  \item[$(i)$] {The size of the search prior is much larger than that of the targeted posterior. As EMRI signals are expected to complete tens of thousands of cycles in the LISA band, parameter measurements are extremely precise (often exceeding one part per million)~\cite{Babak:2017tow,Speri:2021psr,Chapman-Bird:2025xtd}.
  This in turn renders fully-coherent, grid-based matched filter searches~\cite{Usman:2015kfa,Wette:2018bhc,Brady:1998nj,Pletsch:2008gc,Wette:2015lfa,Moore:2019pke}, 
  which would require $\mathcal{O}(10^{40})$~\cite{Gair:2004si} templates, completely unfeasible.}
 
  \item[$(ii)$] {The computational cost of EMRI waveforms is expensive ($\sim 100\, {\rm ms}$ per evaluation) despite substantial recent progress in the form of the FastEMRIWaveforms (\texttt{few}) software package~\cite{Chua:2020stf,Katz:2021yft,Speri:2023jte,Chapman-Bird:2025xtd}. 
  This computational cost is due to both the number of harmonics and the large number of cycles in-band. 
  Due to their small mass ratios, EMRIs evolve slowly and produce long-duration signals with the compact object spending extended periods in the strong-field regime. 
  When combined with orbital eccentricity and orbital-plane precession, this significantly broadens the harmonic spectrum of the GW signal.}

  \item[$(iii)$] {
  The likelihood surface is characterized by many local maxima due to non-local 
  parameter degeneracies~\citep{Chua:2021aah}. %
  While these tertiary maxima are minuscule compared to the global maximum associated with the true source parameters, stochastic search methods can stall at these points in parameter space and fail to locate the global maximum. %
  }
\end{itemize}

To address these challenges, several approaches for EMRI identification have been explored.
Hybrid methods between genetic algorithms and Markov Chain Monte Carlo (MCMC) stochastic searches have been introduced in Refs.~\cite{Strub:2025dfs, Babak:2009ad, Cornish:2008ra}. Using different types of modifications to the search function and/or annealing schemes, the prior-posterior ratio is decreased~\cite{Strub:2025dfs} and using intelligent proposal moves has helped with the local maxima problem~\cite{Babak:2009ad, Cornish:2008ra}.  %
The use of a veto likelihood has also been explored to effectively suppress local maxima~\cite{Chua:2022ssg}. %

Time-frequency methods have been applied to identify EMRI time-frequency tracks on a spectrogram, 
without the need to generate waveforms~\cite{Gair:2007ym,Gair:2006nj,Wen:2006vi}. 
Phenomenological waveforms have been employed to create template families that can cover a large portion
of EMRI parameter space, allowing for model-independent detection before parameter estimation with a 
specific EMRI model~\cite{Wang:2012whc, Ye:2023xka}. 

Machine-learning techniques have also been explored to %
reduce computational costs~\cite{Zhao:2023ncy,Zhang:2022xuq,Yun:2023oqk} and improve sampling efficiency~\cite{Liang:2025vuf}. Sequential simulation-based inference (SBI) methods, specifically Truncated Marginal Neural Ratio Estimation, are being investigated to efficiently narrow down the complex, high-dimensional parameter space, offering a potential solution for EMRI searches and parameter estimation from initially wide priors~\cite{Cole:2025sqo}. Sparse Dictionary Learning (SDL) is being explored for high-speed reconstruction of long-duration EMRI waveforms, demonstrating the potential to significantly accelerate search methods~\cite{Badger:2024rld}.

Despite these advances, a gap remains in the literature: 
the systematic characterization of search performance across the EMRI 
parameter space using single-harmonic detection.
This quantitative assessment is essential for evaluating any search pipeline's sensitivity and reliability. While such characterization may be challenging in realistic LISA data containing multiple overlapping sources, establishing baseline performance for single EMRIs in stationary Gaussian noise remains a fundamental requirement. Key questions that must be addressed include: What signal-to-noise ratio (SNR) threshold enables reliable single-harmonic EMRI detection? How does detection probability vary across the EMRI parameter space, particularly with increasing eccentricity? How accurately can we recover the frequency evolution of the dominant harmonic? Given such a frequency track, how well can we constrain the underlying EMRI parameters?

In this work, we address these questions by developing a search pipeline that takes as input simulated Gaussian noise with an EMRI signal and outputs the best frequency evolution found. %
We adopt a semi-coherent detection statistic, building upon Ref.~\cite{Tenorio:2024jgc},
that lowers the prior-to-posterior ratio at the cost of a loss in 
sensitivity~\cite{Brady:1998nj,Krishnan:2004sv,Pletsch:2008gc,Wette:2015lfa,Tenorio:2021wmz}. 
Template generation is accelerated by representing the EMRI frequency evolution through a Singular Value Decomposition (SVD) basis~\cite{Roulet:2019hzy}. 
We speed up the search statistic computation by adopting the time-frequency domain for a single harmonic waveform and assuming a slowly evolving frequency~\cite{Tenorio:2025gci}.
We implement this in \texttt{jax} to accelerate the computation of the detection statistic and its maximization through automatic differentiation for the stochastic search~\cite{jax2018github}. 
After the frequency evolution is identified, we train a density estimator to find the best EMRI parameters that produce such a frequency track using SBI. 
The key output of our pipeline is a suitable proposal distribution that can be used for follow-up EMRI identification analyses which we explore in Sec.~\ref{subsec:parameter_identification}.
Finally, we discuss the astrophysical implications of this work and its relevance to the LISA Global Fit problem in Sec.~\ref{sec:results}.

\section{Methods}\label{sec:methods}

\subsection{EMRI frequency evolution}\label{subsec:emri_freq_ev}

We consider eccentric equatorial EMRI systems, which consist of the inspiral of a stellar-mass compact object (of mass $m_2$) into a MBH (of mass $m_1$ and dimensionless spin $|a| < 1$). 
{While this represents a subset of general EMRI configurations, we focus on equatorial systems due to the availability of accurate models at adiabatic order. Moreover, since in this work we aim to identify a single harmonic in the data, equatorial systems offer a simpler starting point for our analysis. Generic EMRIs typically exhibit a richer harmonic structure, making them more challenging to analyze.}
The GW signal emitted by the system during inspiral can be expressed (to adiabatic order) as a sum of harmonic modes labeled by pairs of integers \((m, n)\)~\cite{Hughes:2021exa}:
\begin{equation}\label{eq:waveform}
    h_+(t) - i h_\times(t) = \sum_{m,n} A_{m,n}(t) \, e^{i \Phi_{m,n}(t)}\,,
\end{equation}
where the amplitudes \(A_{m,n}(t)\) evolve slowly, on the radiation-reaction timescale, while the phases \(\Phi_{m,n}(t)\) encode the rapid oscillatory behavior of the waveform on the orbital timescale. 
In this work, we focus on the relationship between orbital dynamics and the GW phase evolution of each harmonic, as it determines the frequency evolution of the GW signal at adiabatic order~\cite{Hughes:2021exa}. As we will explore in Section~\ref{subsec:det_stat}, our detection statistic implementation does not require knowledge of \(A_{m,n}(t)\).%

The evolution of $\Phi_{m,n}$ is typically obtained via numerical integration of the equations of motion of the system.
In this work, we use the \texttt{KerrEccEqFlux} model as implemented in the \texttt{few} software package~\cite{Chua:2020stf,Katz:2021yft,Speri:2023jte,Chapman-Bird:2025xtd}.
We briefly summarise this procedure here and refer the reader to Ref.~\cite{Chapman-Bird:2025xtd} for further details.
The inspiral trajectory is described by the evolution of two quasi-Keplerian orbital elements -- semi-latus rectum $p$ and eccentricity $e$ -- evolving under the loss of energy and angular momentum due to GW radiation-reaction.
For convenience, we integrate the trajectory backwards with respect to time from the end of inspiral, which we define by the separatrix $p_{\mathrm{sep}}(a,e_f) + 0.002$ (as specified by the model) and the final eccentricity $e_f$, to a specified time to plunge $T_{\mathrm{pl}}$.
This procedure yields the time evolution of the orbital frequencies $\Omega_\phi(t)$ and $\Omega_r(t)$, from which we compute the GW harmonic frequencies
\begin{align}\label{eq:phase_waveform}
    f_{m,n}(t) \equiv \frac{1}{2\pi} \dv{\Phi_{m,n}}{t} = \frac{m \, \Omega_\Phi(t) + n \, \Omega_r(t)}{2\pi} \, .
\end{align}
Therefore, the harmonic frequency evolution can be determined by the parameters $\theta = \{m_1, m_2, a, T_{\rm pl}, e_f\}\rightarrow \{f_{m,n}(t), \dot{f}_{m,n}(t)\}$ at adiabatic order. This relation will be used to construct a phenomenological model of the EMRI frequency evolution.

The remaining ingredient for the waveform is the generation of amplitudes \(A_{m,n}(t)\) which depend on the orbital elements evolution and the source orientation. We use the \texttt{few} package ~\cite{Chua:2020stf,Katz:2021yft,Speri:2023jte,Chapman-Bird:2025xtd} for the trajectory, amplitude and waveform generation.

\subsection{Semi-coherent detection statistic}\label{subsec:det_stat}
EMRIs are usually characterized by measurement precision of the order of $10^{-5}$ for their intrinsic parameters $m_1, m_2, a, T_{\rm pl}, e_f$ because of their large number of cycles in band,  $\sim 10^5$. This means that the typical prior-to-posterior ratio is $\mathcal{O}( 10^{5})$ for each intrinsic parameter, leading to a total prior-to-posterior ratio of $\mathcal{O}(10^{25})$ which amounts to the number of templates needed.\footnote{Note that Ref.~\cite{Gair:2004si} estimated the number of templates to be $10^{40} = (10^{5})^8$, as they considered generic EMRI systems with 8 parameters.} This renders fully-coherent matched filtering searches unfeasible. Here, we present a semi-coherent detection statistic that reduces the prior-to-posterior ratio at the cost of a loss in sensitivity and we detail how computing this statistic in the time-frequency domain allows for a fast evaluation.

We decompose the GW signal \( h(t; \theta) \) %
into a sum of localized waveforms over short time segments indexed by \( \alpha \), typically corresponding to Short Fourier Transform (SFT) windows~\cite{Tenorio:2025gci}:
\begin{equation}
    h(t; \theta) = \sum_{\alpha} \sft{h}(t; \sft{\theta}) \, .
\end{equation}
Each segment waveform \( \sft{h}(t; \sft{\theta}) \) is non-zero only within its corresponding time window \( \sft{\mathcal{T}} \), i.e., \( \sft{h}(t; \sft{\theta}) = 0 \) for \( t \notin \sft{\mathcal{T}} \). The local parameters \( \sft{\theta} \) are uniquely determined by the global parameters via a mapping \( \sft{\theta} = \sft{\theta}(\theta) \).
This decomposition naturally aligns with the SFT framework~\cite{Tenorio:2025gci},  
where each local waveform can be approximated as a chirping sinusoid:
\begin{align}
    \sft{h}(t; \sft{\theta}) &= \sft{A} \exp\left[i \sft{\phi} + i \Delta \sft{\phi}(t)\right] \, , \\
    \Delta \sft{\phi}(t) &= 2\pi \sft{f} (t - \sft{t}) + \pi \sft{\dot{f}} (t - \sft{t})^2 \, ,
\end{align}
valid within the segment \( t \in \sft{\mathcal{T}} \) for the local parameters \( \sft{\theta} = \{ \sft{A}, \sft{\phi}, \sft{f}, \sft{\dot{f}} \} \). Here, we assumed only one chirping sinusoid per segment, corresponding to the dominant harmonic of the EMRI signal \cite{Candes:2008zx}.

Since we are interested in detecting the signal at the cost of losing sensitivity, we can maximize over $\sft{A}$ and $\sft{\phi}$ parameters in each segment making the search statistic semicoherent~\cite{Prix:2011qv}.
This makes the search less sensitive compared to a fully coherent, matched filtering strategy, 
but also more efficient to compute while decreasing the prior-posterior ratio. 
Such maximizations lead to the following detection statistic:
\begin{equation}\label{eq:det_stat}
    \tilde{\mathcal{S}}(\theta)
    = \sum_\alpha \mathcal{S}_{\alpha}(\sft{f}, \sft{\dot{f}})= \sum_\alpha 
    \frac{\left( \mathrm{max}_{\phi_\alpha} \braket{d_\alpha}{\sft{h}}\right)^2}{\braket{\sft{h}}{\sft{h}}} \,,
\end{equation}
where
\begin{equation}
    \mathrm{max}_{\phi_\alpha}\braket{d_\alpha}{\sft{h}} = 4 \left| \sum_{k=0} ^\infty \frac{\tilde d^* _{\alpha,k} \tilde h_{\alpha,k}}{S_n(f_k)} \Delta  f \right |\,.
\end{equation}
Here \( \tilde{d}_{\alpha,k} \) and \( \tilde{h}_{\alpha,k} \) are the SFTs of the data and template, respectively. The time segment is denoted by \( \alpha \) and the frequency by $k$. The frequency resolution \( \Delta f \) is determined by the time segment duration $\Delta f = 1/T_{\rm SFT}$.
Note that the maximization in each segment over the phase $\phi_\alpha$ is equivalent to taking the absolute value instead of the real part of the integral.
We have now reduced the local parameters to \( \sft{\theta} = \{\sft{f}, \sft{\dot{f}} \} \) at the cost of maximizing over two parameters per segment.

We evaluate the detection statistic \( \tilde{\mathcal{S}} \) using the linear chirp approximation introduced in Ref.~\cite{Tenorio:2025gci}, which enables efficient inner product calculations directly in the SFT domain~\cite{Cornish:2020odn}.
The linear chirp approximation increases the efficiency of the inner product calculation by reducing 
the number of frequency bins to integrate in the inner product from $N_\alpha \times N_f$ ($N_{f} \sim 10^4$)
to $N_\alpha \times (2 P+1)$, where $P$ corresponds to the typical bandwidth of a signal (in bins) within
an SFT. We take $P=100$ since the frequency derivative is in the range $\log_{10}\dot f \in [-12,-8]$,
well within the ranges discussed in Ref.~\cite{Tenorio:2025gci}.

We adopt a segment duration of \( T_{\rm SFT} = 5 \times 10^4  \)~s, which balances several considerations. The resulting time segments are
as long as possible, to minimize the impact on noise due to the use of a semicoherent statistic,
but short enough for the linear approximation to remain valid across a wide range of EMRI signals. 
This approximation breaks down near plunge, where the signal evolution becomes highly nonlinear. To ensure validity, we exclude segments where the second-order frequency evolution exceeds the threshold discussed
in Ref.~\cite{Tenorio:2025gci}: 
$\tilde{\mathcal{S}}_\alpha = 0 \quad \text{if} \quad \ddot{f}_\alpha T_{\rm SFT}^3 / 6 > 1 \,$.
Across the EMRI parameter space considered, we find that for 
\mbox{$T_{\rm SFT} = 5 \times 10^4\,\mathrm{s}$},
99.6\% of segments satisfy \( \ddot{f}_\alpha T_{\rm SFT}^3 / 6 < 1 \), resulting in 631 valid SFT segments.
Increasing \( T_{\rm SFT} \) by a factor of 5 reduces coverage to 80.8\%, with only 126 valid segments. Reducing \( T_{\rm SFT} \) to \( 1 \times 10^4 \) s ensures all 3155 segments satisfy the above condition, significantly increasing the computational cost of the analysis.
The study of other choices of $T_{\rm SFT}$ and their impact on the detection statistic is left for future work.

\subsection{Singular Value Decomposition basis for EMRI frequency evolution}

Next, let us address how to efficiently obtain the parameters 
$\sft{\theta} = \{\sft{f}, \sft{\dot{f}} \}$ necessary for evaluating the detection statistic.
The frequency evolution of EMRIs is determined by the global parameters $\theta = \{m_1, m_2, a, T_{\rm pl}, e_f\}$ through the trajectory evolution:
\begin{equation}\label{eq:theta_traj_f}
    \theta = \{m_1,m_2,a,T_{\rm pl}, e_f\} \xrightarrow{\mathrm{Trajectory}}  f_{m,n}(\sft{t}; \theta)\,,
\end{equation}
where $\xrightarrow{\mathrm{Trajectory}}$ indicates the solution of the trajectory differential equations described in Sec.~\ref{subsec:emri_freq_ev}.
The computational cost of evaluating the trajectory is about $10-50$~ms in FEW~\cite{Chapman-Bird:2025xtd}, and searching directly in the EMRI parameter space $\theta$ is impractical due to strong correlations~\cite{Chua:2021aah}.
For these reasons, previous studies have employed phenomenological models based on Taylor expansions of the frequency evolution~\cite{Wang:2012whc,Ye:2023xka}. Here, we explore a principal-component approach, inspired by Ref.~\cite{Roulet:2019hzy}, where the frequency evolution is represented as a linear combination of basis functions with associated coefficients. This makes the determination of the frequency evolution more efficient and reduces correlations in the search space. We now detail how this basis is constructed. 

We assume that the dominant harmonic is given by the $m=2$ mode and drop the index for ease of notation, i.e. $f = f_{m=2,n=0}$.  This assumption is valid in the low-eccentricity limit, where the majority of the emitted signal power is in the dominant mode. %
Therefore, we use a flat prior in final eccentricity up to $e_f=0.1$ for our search parameter space shown in Table~\ref{tab:emri_priors}.
The prior on the primary mass is uniform in logarithmic space, which is a reasonable choice given the expected astrophysical population of EMRIs~\cite{Lyu:2024gnk}. The choice of uniform priors in the secondary mass and primary spin is somewhat arbitrary given the current uncertainties in EMRI formation channels. We will revisit this in future work and discuss its impact below.
\begin{table}
\centering
\begin{tabular}{l@{\hskip 0.1in}|@{\hskip 0.1in}c@{\hskip 0.1in}|@{\hskip 0.1in}c}
\textbf{Parameter} & \textbf{Symbol} & \textbf{Prior Range} \\
\hline \hline
Primary mass & $m_1$ & \( [0.5, 5 ] \times 10^6 M_\odot \) \\
Secondary mass & \( m_2 \) & \( [1, 100] \, M_\odot \) \\
Primary spin & \( a \) & \( [0, 0.998] \) \\
Plunge time & \( T_{\rm pl} \) & \( [0.1, 1.0] \, \text{yr} \) \\
Final eccentricity & \( e_f \) & \( [0, 0.1] \) \\
\end{tabular}
\caption{
Prior ranges for the EMRI intrinsic parameters used in the search. Priors are uniform in all parameters apart from $m_1$ where it is uniform in $\log_{10} m_1$.
}
\label{tab:emri_priors}
\end{table}

We generate $10^4$ realizations of EMRI parameters using Eq.~\eqref{eq:theta_traj_f} to obtain the frequency evolution $f(\sft{t}; \theta_i)$  for a fixed time to plunge $T_{\rm pl}=1$ year.
We use the SVD to find a representation of the frequency evolution in a similar way to~\cite{Roulet:2019hzy}:
\begin{equation}
    f(\sft{t}; \theta_i) = f_{i,\alpha} = \sum_p U_{i,p} S_{p}V_{p,\alpha}\,.
\end{equation}
We find that, for the priors considered here and fixed time to plunge, we need only up to 20 singular values $S_p$ to obtain a relative precision of $10^{-7}$ in the reconstructed $f_{i,\alpha}$. We can therefore search for a frequency evolution in the parameter space of $\xi_p$ where:
  \begin{equation}
      f_\alpha \approx \sum^{20} _{p=1} \xi_p V_{p,\alpha}\,.
\end{equation}

This procedure has three key advantages. First, it projects the high-dimensional frequency-evolution space into a compact representation that preserves most information across broad EMRI parameter priors. 
Second, it eliminates the need for computationally expensive trajectory evaluations, instead enabling efficient frequency generation through matrix multiplication. Third, it allows computation of the detection statistic $\tilde{\mathcal{S}}(\sft{f}, \sft{\dot f})$ in terms of the coefficients $\xi_p$ in a differentiable manner, enabling the gradient-based optimization strategy described in the next section and implemented in \texttt{jax}. The frequency derivative $\dot f_\alpha$ is computed numerically from the frequency evolution $f_\alpha$. This is sufficient to achieve a 0.5\% (1\%) accuracy in $\dot f$ up to $10 $ $(5)$ days from plunge.

The downside is that we must define priors for the coefficients $\xi_p$, we still need to include the time to plunge parameter, and we lose the direct connection to the physical EMRI parameters $\theta$. We address the first issue here and defer the discussion of the latter two to Secs.~\ref{subsec:search} and \ref{subsec:sbi}, respectively.
We define a uniform prior with boundaries based on the symmetric 3$\sigma$ quantiles of the coefficients obtained with $U_{i,p} S_{p}$. The uniform prior is justified by the fact that the Euclidean distance in this space coincides with the mismatch distance between similar waveforms. One can see this by considering two nearby templates $h_\alpha (\xi_p)$ and $h_\alpha (\xi_p ')$ where their mismatch is proportional to $|f_\alpha - f_{\alpha}'|^2 \sim |\xi_p - \xi_p'|^2$. Therefore, these coordinates $\xi_p$  are naturally suited for geometric placement of templates and for search optimization~\cite{Roulet:2019hzy}. 
\begin{figure}
  \centering
  \includegraphics[width=\columnwidth]{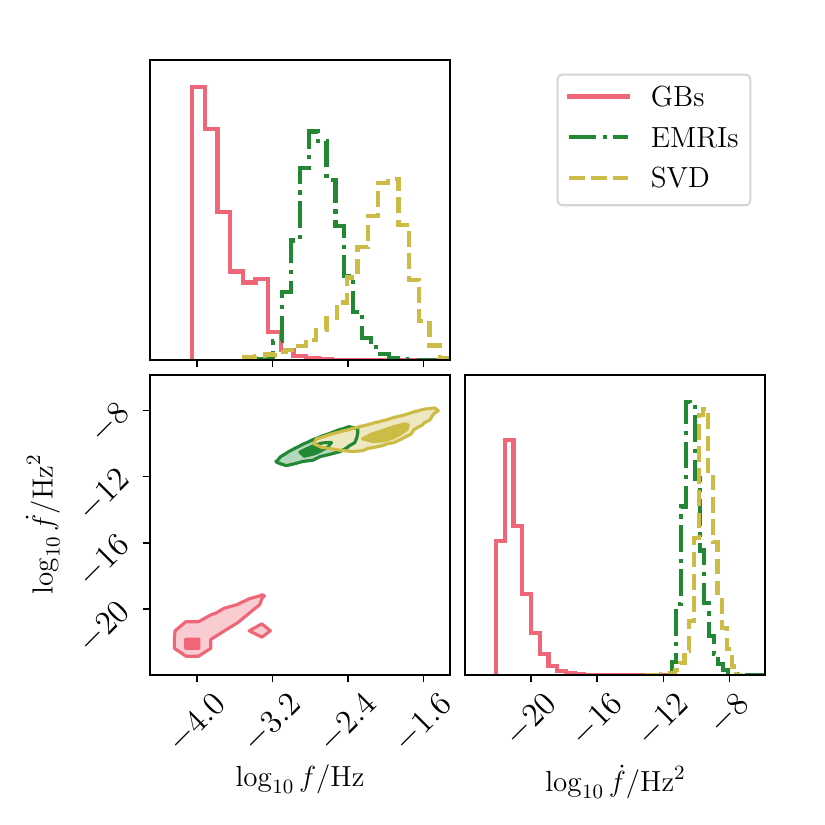}
  \caption{
 Frequency $f$ and frequency derivative $\dot{f}$ distributions for galactic binaries (GBs; pink, solid histogram), EMRIs (green, dashed-dotted histogram), and EMRI phenomenological frequency evolution constructed using the SVD basis (yellow, dashed histograms). The galactic-binary sample is taken from one of the LISA Data Challenges~\cite{Baghi:2022ucj}. 
 EMRI samples represent the full frequency evolution of the dominant azimuthal mode $m=2, n=0$ across the searched parameter space of Table~\ref{tab:emri_priors}. The SVD tracks are obtained by uniformly sampling the SVD coefficient space $\xi_p$ and reconstructing the frequency tracks. These two different priors (EMRIs and SVD) lead to two different frequency and frequency derivative densities.
 Contours mark 68\% and 95\% credible regions. 
  }
  \label{fig:frequency_prior}
\end{figure}
We verified that such a prior covers most of the frequency tracks in the EMRI parameter space of Table~\ref{tab:emri_priors}.

Figure~\ref{fig:frequency_prior} shows the frequency $f_\alpha$ and frequency derivative $\dot f_\alpha$  distributions obtained by uniformly sampling in either the EMRI parameter space or the SVD basis prior. A uniform prior in the SVD basis leads to higher frequency derivatives with respect to the EMRI basis. This is expected because the SVD basis places more templates for sources with larger number of cycles $N_{\rm cyc}\sim f_\alpha^2/\dot f_\alpha$. For reference, we also show the distribution of galactic binaries from one of the LISA Data Challenges~\cite{Baghi:2022ucj}. This is useful to show that the EMRIs and the galactic binaries have distinct frequency evolution characteristics.

\subsection{Search strategy}\label{subsec:search}

Here, we define the search strategy to find the optimal SVD basis coefficients $\xi_p$ and time to plunge $T_{\rm pl}$ that maximize the detection statistic. Since our SVD basis is constructed for a fixed plunge time of 1 year, we must search over the unknown actual plunge time $T_{\rm pl}$. 

To this end, we analyze data segments of progressively increasing duration $T_{\rm obs}$ and design the search such that the plunge occurs at the end of each segment ($T_{\rm pl} = T_{\rm obs}$). For each assumed plunge time, we apply an integer time shift to the pre-computed SVD basis to align the frequency track accordingly. This shifting strategy avoids the computational cost of reconstructing the SVD basis for each $T_{\rm pl}$ value and maintains differentiability for gradient-based optimization with respect to $\xi_p$.

We search over a year-long dataset, consistent with the expected duration of the first LISA data release~\cite{LISA:2024hlh}. We analyze progressively longer data streams starting from a minimum duration of 0.1~yr in increments of approximately 7~days, corresponding to 12~SFTs. For each observed data chunk of duration $T_{\rm obs}$, and therefore for each considered $T_{\rm pl}$, we search for the best-fit parameters $\xi_p$. The maximization is performed using a stochastic search combining gradient-based and population-based methods, as detailed below. This search strategy provides us with 48 best fit parameters $(\xi_p,T_{\rm pl})$ that can be used to define a detection statistic as a function of observation time.
We also enforce that there are no contributions to the detection statistic for SFTs with frequencies below $10^{-3.5}\approx 3\times 10^{-4}$ Hz and frequency derivatives below $\dot f < 10^{-13} $ Hz/s. 

The maximization procedure begins by generating an initial population of 1024 candidate solutions randomly sampling the SVD coefficients $\xi_p$ from uniform priors derived from the EMRI parameter boundaries defined in the previous section. %
To improve coverage across the relevant frequency range, we enforce that each track's initial frequency lies between $5 \times 10^{-4}$ Hz and $10^{-2}$ Hz by uniformly sampling this range and projecting the resulting frequency onto the SVD basis. After evaluating the detection statistic for all candidates, we retain the 512 highest-scoring solutions to seed the optimization. This reduction balances the need for high-quality initial conditions with the GPU memory constraints of the subsequent iterative procedure.

The optimization combines gradient-based and population-based methods, leveraging \texttt{jax} for efficient computation. Specifically, we perform 525 iterations where, at each step, we perform two optimizations. An Adam optimizer (learning rate $10^{-5}$) is first applied to refine the $\xi_p$ values locally, with gradients computed via automatic differentiation with the package \texttt{optax}~\cite{deepmind2020jax}. 
This is followed by a differential evolution (DE) step to promote global exploration and escape local minima. 
The large number of walkers (512) enables broad exploration of the parameter space, 
with Adam providing efficient local descent throughout the 525 optimization steps. We verified that our results do not vary with an increasing number of optimization steps. Similar optimization techniques have been explored in Ref.~\cite{Green:2024yka}.

The DE step is performed in the frequency-track space as follows. The current population of $\xi_p$ is transformed to $f_\alpha$ tracks, a DE trial vector is generated for each walker according to:
\begin{equation}\label{eq:de_step}
    f_\alpha ^{\mathrm{trial}} = f_\alpha + F_{\rm DE} (f_{\alpha,r_1} - f_{\alpha,r_2}) \,,
\end{equation}
where $r_1$ and $r_2$ are two randomly selected walkers different from the current population, and $F_{\rm DE}$ is the differential weight randomly drawn between 0.3 and 1.5. If the trial vector improves the detection statistic and it is within the prior boundaries, the trial candidates $f_\alpha$ are projected back to $\xi_p$ via the pseudo-inverse of the SVD basis and are accepted.
We {empirically} found this track-space DE to be more efficient than applying DE directly in $\xi_p$ space because it allows walkers stuck on different harmonics to jump between them, similarly to the MCMC jumps of Ref.~\cite{Cornish:2008ra}. 
Note that the DE step of Eq.~(\ref{eq:de_step}) does not include crossover in order to ensure the continuity of the trial solution.

Since the search is performed over progressively longer data segments, the detection statistic becomes a function of the observed duration. As expected, both the mean and standard deviation of the detection statistic increase with longer observation times due to the growing number of amplitude and phase maximizations. To account for this, we normalize the detection statistic $\tilde{\mathcal{S}}$ for each data duration using its empirically estimated mean and standard deviation under noise-only injections. This normalization is applied consistently when analyzing both noise-only and signal-injected data. We stress that this does not ensure that they share the same statistical properties,
but it is a practical choice for finding the best fit parameters across all tested durations, i.e.,
across the 48 best fit parameters $(\xi_p,T_{\rm pl})$. In the remaining, we refer to the normalized detection statistic simply as \( \mathcal{S} \).

\begin{figure}
  \centering
  \includegraphics[width=\columnwidth]{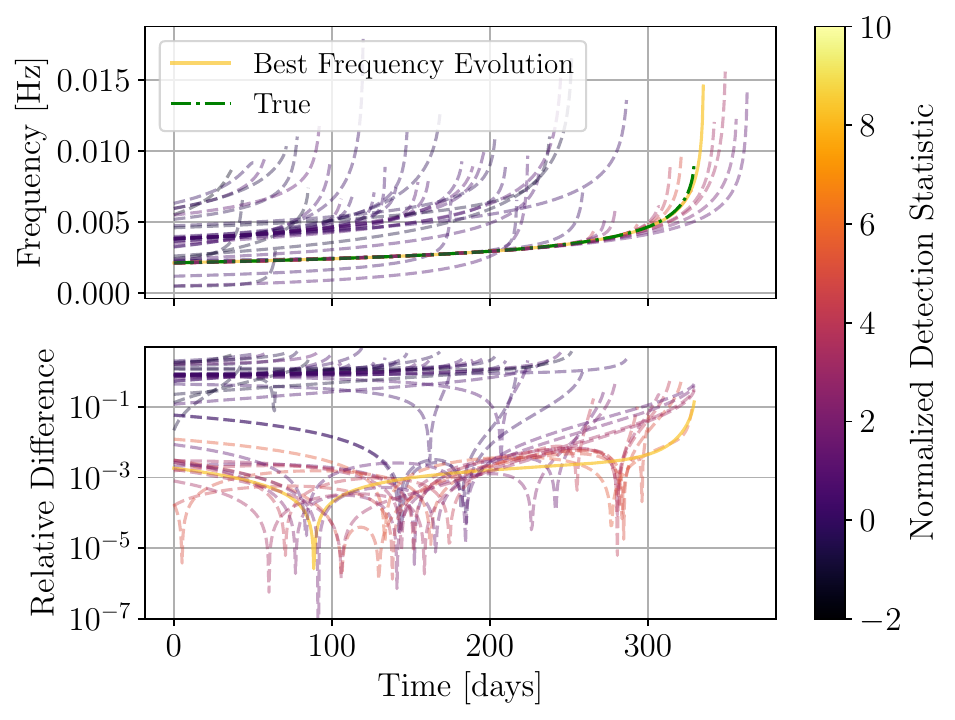}
  \caption{
    Recovered frequency evolution of the search of an EMRI injection with primary mass $m_1 = 1.3379 \times 10^6\,M_\odot$, secondary mass $m_2 = 27.091\,M_\odot$, primary spin $a = 0.8636$, time to plunge $T_{\rm pl} = 0.902\ \mathrm{yr}$, and final eccentricity $e_f = 0.007086$, and $\mathrm{SNR} = 30$. \textit{Top panel:} Best fit frequency evolution as a function of different observed data durations, color coded by the normalized detection statistic. The dashed lines show the best frequency evolutions obtained per each observed data duration.
    The dashed-dotted green line represents the true injected frequency track from the dominant $m=2, n=0$ harmonic, whereas the solid yellow line represents the best frequency evolution obtained by the search strategy across all observed data durations. The search of the entire one year dataset takes 1.1 hours and evaluates $\approx 10^7$ detection statistics. \textit{Bottom panel:} The relative difference $|\Delta f/f|$ between the recovered and true frequency tracks as a function of time.
  }
  \label{fig:frequency_recovery}
\end{figure}

An example of the best fit frequencies obtained through this process for different observation durations is shown in Fig.~\ref{fig:frequency_recovery}.
It can be seen that different observation durations lead to different best-fit frequency tracks ending at different times defined by the increments of the search strategy. As the observation duration gets closer to the true plunge time, the normalized detection statistic increases and the relative difference between the recovered and true frequency tracks decreases. The total number of templates $\approx 10^7$ used in the search is given by the number of walkers active in the stochastic optimization (512), times the number of maximum optimization steps (525) and the number of data chunks analyzed (48). The total run time for the analysis is of 1.1 hours on 1 GPU.

\subsection{Simulation based inference for the EMRI inverse problem}\label{subsec:sbi}
Once a candidate frequency evolution $f_\alpha$ is identified by the search pipeline, we are interested in using the recovered frequency track to find the underlying EMRI parameters.
This can be considered an inverse problem where we want to estimate the EMRI parameters $\theta$ given the recovered frequency evolution $f_\alpha^{\rm rec} \rightarrow \theta$.
This can be done by minimizing the difference between the recovered frequency evolution and the frequency evolution predicted by the EMRI model for different parameter values as done in~\cite{Cornish:2008ra}.
This approach allows us to leverage the information contained in the frequency evolution to constrain the EMRI parameters $\theta$. 

However, since the recovered frequency evolution is subject to uncertainties, multiple EMRI parameters can give rise to very similar frequency tracks~\cite{Chua:2021aah}. We must account for these factors in our estimation process. This can be achieved by incorporating a Bayesian framework, where we treat the EMRI parameters as random variables and update our beliefs about their values based on the observed data. To do so we need to estimate the uncertainty on the recovered frequency evolution. This can be done by analyzing the distribution of recovered frequency evolutions from multiple searches. If we have an estimate of the uncertainty $\sigma$ with which our pipeline recovers on average the frequency evolution, we can define a data generation process $f_\alpha^{\rm rec} = f_\alpha(\theta)(1 + \epsilon_\alpha)$, where $\epsilon_\alpha$ is a Gaussian noise term with zero mean and standard deviation $\sigma$. 
The noise characteristics in the recovered frequency evolution are generally unknown and may deviate from Gaussianity. Moreover, the recovered signal may not align perfectly with the EMRI model, as it is obtained via a phenomenological SVD basis fit. Nonetheless, assuming Gaussian noise provides a practical starting point for the inverse problem and remains valid as long as the errors are sufficiently large to avoid biasing the mapping.

We could either sample the distribution of the EMRI parameters using MCMC sampling methods or we could use simulation based inference to directly learn the mapping from the recovered frequency evolution to the EMRI parameters. In~\cite{Wang:2012whc,Ye:2023xka} the authors used an MCMC method to sample the distribution of the EMRI parameters given the recovered frequency evolution.
We found that standard proposals and MCMC samplers such as \texttt{eryn}~\cite{Karnesis:2023ras} struggle to sample the likelihood due to the strong correlations. We therefore propose to use simulation based inference conditioned on the recovered frequency evolution to efficiently sample the distribution of the EMRI parameters, a similar approach was explored in Ref.~\cite{sbi_cw_LVK} for continuous wave searches. Since we are interested in only one recovered frequency evolution, sampling from the prior can be inefficient in the number of simulations, because one is effectively learning a density estimator for all observations in the prior space.
Multi-round inference begins by sampling from the prior, simulating data, and training a neural network to estimate the parameter distribution. It then refines this estimate over multiple rounds, each time conditioning on the recovered frequency track observation. In each round, samples are drawn from the updated distribution, simulated, and used to retrain the network. This iterative process improves the posterior approximation but sacrifices amortization.

This approach allows us to directly learn the mapping from the recovered frequency evolution to the EMRI parameters, bypassing the need for explicit likelihood evaluations and assuming a frequency evolution accuracy $\sigma$.
By leveraging simulation based inference, we can efficiently explore the parameter space and obtain a proposal distribution that could inform subsequent follow-up analyses.

The implementation makes use of the \texttt{sbi} package~\cite{BoeltsDeistler_sbi_2025}.
We define a simulator function that generates synthetic frequency tracks and their derivatives from EMRI parameters $\theta = \{\log_{10} m_1, m_2, a, T_{\rm pl}, e_f\}$ drawn from uniform priors. Gaussian noise with standard deviation $\sigma = 10^{-2}$.
The training proceeds through multiple rounds: $(i)$ sample parameters from the current proposal distribution, $(ii)$ simulate corresponding frequency tracks using the EMRI waveform model, 
$(iii)$ train a normalizing flow to estimate $p(\theta | \{f_{\alpha}\})$ on the simulated data, and $(iv)$ update the proposal distribution using the trained distribution conditioned on the observed track $f_\alpha^{\rm rec}$. This sequential approach progressively focuses sampling around the distribution, improving efficiency compared to sampling from broad priors.
The neural density estimator uses a neural spline flow with parameter-wise normalization to handle the multi-scale nature of EMRI parameters. After convergence, the samples provide parameter estimates with quantified uncertainties, enabling direct assessment of what EMRI configurations are consistent with the recovered frequency evolution.

\subsection{Data generating process}\label{subsec:data_generation}
We define the data generating process for testing the search pipeline.
We assume that the data is composed of a single EMRI signal in instrumental Gaussian stationary noise. In reality, LISA data is expected to contain various source types and noise artifacts which will need to be fitted simultaneously~\cite{Deng:2025wgk,Katz:2024oqg,Strub:2024kbe,Littenberg:2020bxy,Littenberg:2023xpl,Speri:2022kaq}. %
In this context, we assume that an initial step of noise characterization and source subtraction has already been performed.
Even though the LISA global fit is the realistic scenario for an EMRI search, searching for a single EMRI is still an open and challenging problem. %
Recent work~\cite{Khukhlaev:2025xiz} showed that the resolvable galactic binaries strongly contaminate the detection and inference of EMRIs. 
For reference, we show in Fig.~\ref{fig:frequency_prior} the parameter space of frequencies and frequency derivatives for EMRIs and galactic binaries from the Sangria LISA Data Challenge Dataset~\cite{Baghi:2022ucj}.
Future work will need to investigate how our results change in the presence of overlapping sources and find a way to exploit the separation of these two source types in the frequency and frequency derivative parameter space shown in Fig.~\ref{fig:frequency_prior}. 

The signal is generated using \texttt{few}~\cite{Chua:2020stf,Katz:2021yft,Speri:2023jte,Chapman-Bird:2025xtd} assuming Kerr equatorial orbits, resulting in the two polarizations $h_+(t)$ and $h_\times(t)$. This model is the current state-of-the-art approximant for EMRI waveforms and we refer the reader to Ref.~\cite{Chapman-Bird:2025xtd} for a detailed description.
The EMRI parameters are drawn from the priors listed in Table~\ref{tab:emri_priors}. The sky and orientation angles are uniform on the sphere, and the initial phases uniform in $[0, 2\pi)$. The final signal is then obtained after applying the LISA detector response to the two polarizations to compute the relevant second generation time-delay interferometry (TDI) variables using \texttt{fastlisaresponse}~\cite{michael_katz_2025_17162632,Katz:2022yqe}. The luminosity distance is fixed based on the targeted SNR. 
For simplicity, we only use the A TDI channel~\cite{1999ApJ...527..814A}; this choice has the practical benefit of reducing the memory footprint on GPUs. 

The motion of the LISA constellation induces Doppler modulations in the observed frequency evolution $f_{o}$ of GW sources inducing a frequency relative change $f_{o}/f-1\sim v/c$ proportional to the spacecraft velocity $v$ in units of the speed of light $c$.
The motion of the constellation center around the Sun results in velocities of the order of $v \sim 10^{-4}c$~\cite{Marsat:2018oam}. Additionally, the cartwheeling motion of the constellation contributes in velocities of the order of $v \sim 1.7 \times 10^{-6} c$~\cite{Marsat:2018oam}.
These effects induce a phase shift at most of the order of $\Delta \phi \sim f T_{\rm SFT} (v/c)\sim 0.05$ radians over the SFT window $T_{\rm SFT}=5\times 10^{4}$ for a frequency of $10^{-2}$ Hz. Therefore, the phase modulations induced by the LISA motion are relatively small over the SFT duration and can be neglected in our analysis~\cite{Jaranowski:1998qm}.

Noise is generated based on the LISA TDI A sensitivity curve~\cite{Babak:2021mhe}, assuming a known power spectral density (PSD). In principle, we will not have access to the PSD and we will need an estimate from the global fit pipeline. Here, we assume to know the PSD. 
The signal and noise data are generated at a sampling rate of 5~s and for a total duration of 1~yr. 
The time series containing the signal plus noise is then converted in the time-frequency domain using a SFT with time segment duration $\Tsft=5\times 10^4~{\rm s}\approx 14$ hr.

\subsection{Detection probability and false alarm rate}
We quantify the performance of our search strategy by estimating its sensitivity in terms of the detection probability $p_{\rm det}$ and false-alarm probability $p_{\rm FA}$. %
The detection probability represents the fraction of datasets for which the search pipeline produces a detection statistic above a predefined threshold $\mathcal{S}_{\rm th}$.
Let us indicate with $\Pi$ the set of assumptions about the EMRI signal model and noise properties (Sec.~\ref{subsec:data_generation}) and search strategy (Sec.~\ref{subsec:search}). We can then define the probability $p(\mathcal{S} \mid \mathrm{SNR}, \Pi)$ to obtain a detection statistic $\mathcal{S}$ after applying the search strategy to a set of injections according to assumptions $\Pi$ with a given EMRI signal SNR~\cite{Tenorio:2024jgc,Searle:2008jv}. This quantity $p(\mathcal{S} | \mathrm{SNR}, \Pi)$ can be obtained empirically by analyzing multiple realizations of the data generating process and computing the distribution of the resulting detection statistics obtained through the search strategy.

Then, the false-alarm probability for a given threshold $\mathcal{S}_{\rm th}$ is obtained by the fraction of realizations under the null hypothesis (noise only, $\mathrm{SNR}=0$) that exceed the threshold:
\begin{equation}
    p_{\rm FA}(\mathcal{S}_{\rm th}) = p(\mathcal{S} > \mathcal{S}_{\rm th}| \mathrm{SNR}=0, \Pi) \, .
\end{equation}
To extrapolate the false-alarm probability to low values, we fit the distribution $p(\mathcal{S} | \mathrm{SNR}, \Pi)$ with a Gumbel distribution as in Ref.~\cite{Tenorio:2021wad}.
Given a fixed false-alarm probability \( p_{\rm FA} \), the detection probability for signals with SNR is defined as:
\begin{equation}
    p_{\rm det}({\rm SNR}, p_{\rm FA}) = p(\mathcal{S} > \mathcal{S}_{\rm th}(p_{\rm FA}) | \mathrm{SNR}, \Pi) \, .
\end{equation}
Since we always apply the same search strategy and data generation assumptions in the remaining of this work, we drop the dependency on $\Pi$ from the notation $p_{\rm det}$ and $p_{\rm FA}$.

\section{Results}\label{sec:results}

\subsection{Search under noise and signal hypothesis}

\begin{figure}
  \centering
  \includegraphics[width=\columnwidth]{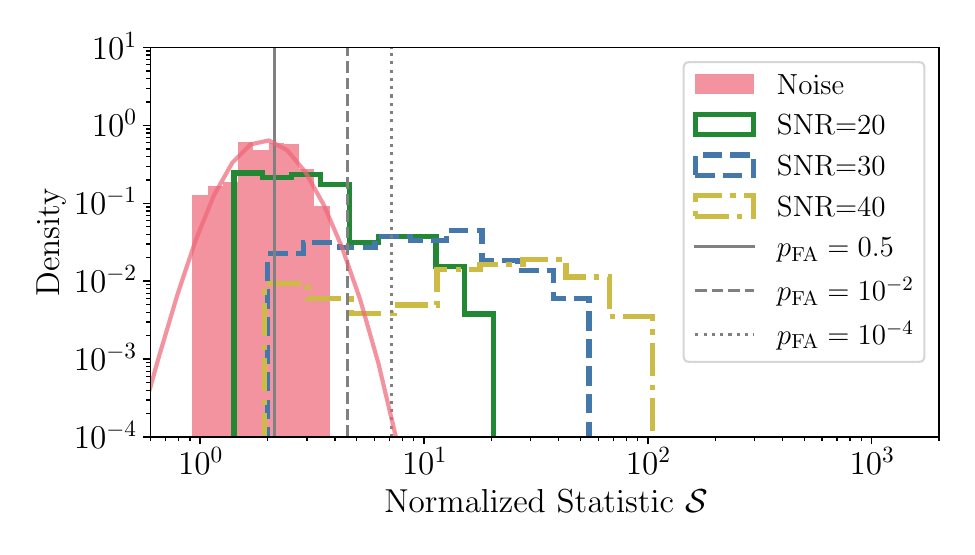}
  \caption{
  Distribution of normalized detection statistic $\mathcal{S}$ under the noise-only hypothesis (pink) and signal+noise hypotheses at different SNRs. Each histogram is constructed from 100 data realizations (Sec.~\ref{subsec:data_generation}).
  The pink noise-only histogram establishes the detection threshold at three different false-alarm probabilities $p_{\rm FA}=0.5$ (solid grey), $p_{\rm FA}=0.01$ (dashed grey), and $p_{\rm FA}=10^{-4}$ (dotted grey). A fit to the noise distribution is shown as solid line.
  The overlaid empty histograms show results from EMRI signal injections with SNR = 20 (solid green), SNR = 30 (dashed blue), and SNR = 40 (dashed-dotted yellow).
  }
  \label{fig:score_histogram}
\end{figure}

To evaluate the performance of our search pipeline, we characterize its behavior under both the null hypothesis (data consisting solely of noise) and the alternative hypothesis (data containing an injected EMRI signal). 
The datasets are generated as described in Sec.~\ref{subsec:data_generation} for SNR$=0, 20, 25, 30, 35,$ and $40$. For each of these SNRs, we generate 100 independent realizations of the data and run the search pipeline as described in Sec.~\ref{subsec:search}. Each search returns a maximum normalised detection statistic $\mathcal{S}$ and the corresponding best-fit frequency evolution $f_\alpha^{\rm rec}$.%

The distribution of the detection statistic from the noise-only searches $p(\mathcal{S}| \mathrm{SNR}=0)$ is shown in Fig.~\ref{fig:score_histogram}. Because our search strategy involves maximizing the detection statistic, the distribution of its maximum under the noise hypothesis follows a Gumbel distribution, as established via extreme value theory in Ref.~\cite{Tenorio:2021wad,Croce:2004pv}. 
Accordingly, we fit a Gumbel distribution (pink solid line) and obtain a location parameter $1.94$ and scale parameter $0.56$ using \texttt{scipy}~\cite{2020SciPy-NMeth}. 
This fit allows us to determine detectability thresholds corresponding to false alarm probabilities $p_{\rm FA}=0.5, 0.01,$ or $10^{-4}$ indicated by vertical solid, dashed and dotted lines respectively.

The distributions of the detection statistic under the signal hypothesis are shown in Fig.~\ref{fig:score_histogram} for SNR~$=20, 30,$ and $40$.
The spread in the distributions is due to the different parameters and noise realizations and spans approximately an order of magnitude in $\mathcal{S}$. The median of the distributions increases with the SNR, as expected, with median values of $\mathcal{S}\simeq 5, 10, 20$ for SNR~$= 25, 30, 35$, respectively. For the highest value SNR~$=40$ considered here, the distribution still presents some candidates below $p_{\rm FA}=0.01$.

\subsection{Detection probability}

We quantify the overall sensitivity of our method by computing the detection probability as a function of SNR spanning the parameter prior reported in Table~\ref{tab:emri_priors}. A detection is declared if the normalized statistic {$\mathcal{S}$} exceeds a threshold corresponding to a given false-alarm probability. The latter is determined from the noise-only distribution as shown in Fig.~\ref{fig:score_histogram} and described in the previous section. We choose detectability thresholds
corresponding to $p_{\rm FA}=0.5, 0.01,$ or $10^{-4}$ spanning from optimistic to conservative; similar use cases can be found in the literature~\cite{Wette:2021tbv, Tenorio:2021wad}.

\begin{figure}
  \centering
  \includegraphics[width=\columnwidth]{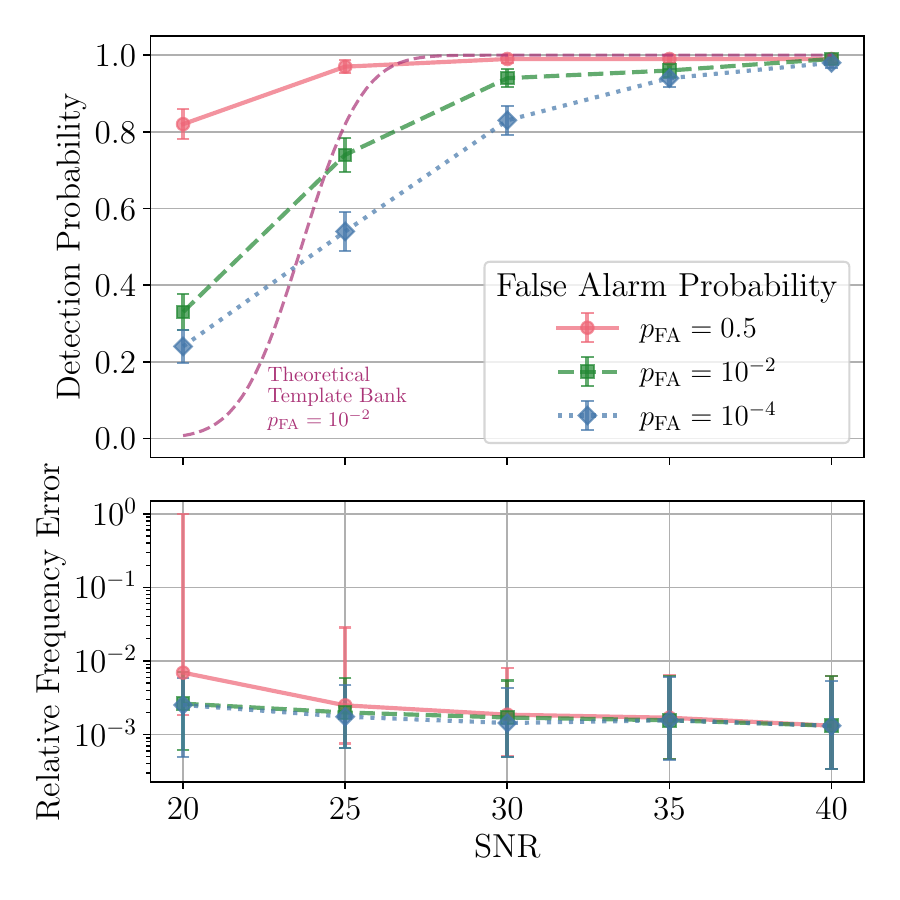}
  \caption{Detection probability and frequency reconstruction accuracy as functions of the signal-to-noise ratio (SNR). \textit{Top panel:} Detection probability curves for false alarm rate thresholds $p_{\rm FA}$: $0.5$ (pink circles), $10^{-2}$ (green squares) and $10^{-4}$ (blue diamonds). Each curve is computed from 100 independent EMRI injections with parameters uniformly sampled across the full-search prior space of Table~\ref{tab:emri_priors}.
  \textit{Bottom panel:} Median relative frequency error $|\delta f/f|$ of recovered frequency tracks from the search pipeline and 1$\sigma$ error bars representing the standard deviation across injections.
  }
  \label{fig:det_prob}
\end{figure}

The top panel of 
Fig.~\ref{fig:det_prob} shows the detection probability as a function of the SNR for representative false alarm probabilities, 0.5, $10^{-2}$ and $10^{-4}$. At a false-alarm probability  $p_{\rm FA}= 0.01$, the detection probability is $\sim74\%$ at SNR~$=25$, $\sim 94\%$ at SNR~$=30$, and $\sim 99\%$ at SNR~$=40$.
At a false-alarm probability of $p_{\rm FA}= 0.5$, the detection probability for SNR$\geq 25$ is above $90\%$.
These results show that searching for a single harmonic is sufficient to identify more than 90\% of the injected EMRI sources with SNR~$=35$ {considered} at a false-alarm probability of 0.01.

The false-alarm probability is here defined for the entire search and cannot be easily compared with the per-template false-alarm probability used in previous works, see e.g. Ref.~\cite{Gair:2004si,Chua:2017ujo,Jaranowski:2005hz}. For a qualitative comparison, Fig.~\ref{fig:det_prob} shows the theoretical, template-bank detection probability for a semi-coherent search.
This is computed using Eq.~(71) of Ref.~\cite{Chua:2017ujo} with $N_\alpha=631$ (the number of SFTs used in this paper) and a per-template false-alarm probability of $10^{-2}/10^{25}$, where $10^{25}$ is the estimated number of independent templates considered in our parameter space and $10^{-2}$ is the overall false-alarm probability $p_{\rm FA}$. This is different from our strategy, where the search is a maximization process and not a template bank search.

The lower panel of Fig.~\ref{fig:det_prob} shows the median relative frequency error ($ |\delta f / f| $) of recovered tracks as a function of the SNR. The error decreases from $\sim 10^{-2}$ at SNR~$=20$ to $\sim 10^{-3}$ at SNR~$=40$, indicating that higher SNRs not only improve detection probabilities but also enhance the accuracy of frequency-track reconstruction. These results are consistent with the search example of Fig.~\ref{fig:frequency_recovery}, where the recovered best-fit frequency track has a median relative error above $\sim 10^{-3}$ at SNR=30. Previous works found relative precision in the recovered frequencies of the order of $10^{-3}$~\cite{Wang:2012whc} and $10^{-4}$~\cite{Ye:2023xka}.

\subsection{Search sensitivity across parameter space}

\begin{figure*}
  \centering
  \includegraphics[width=1.9\columnwidth]{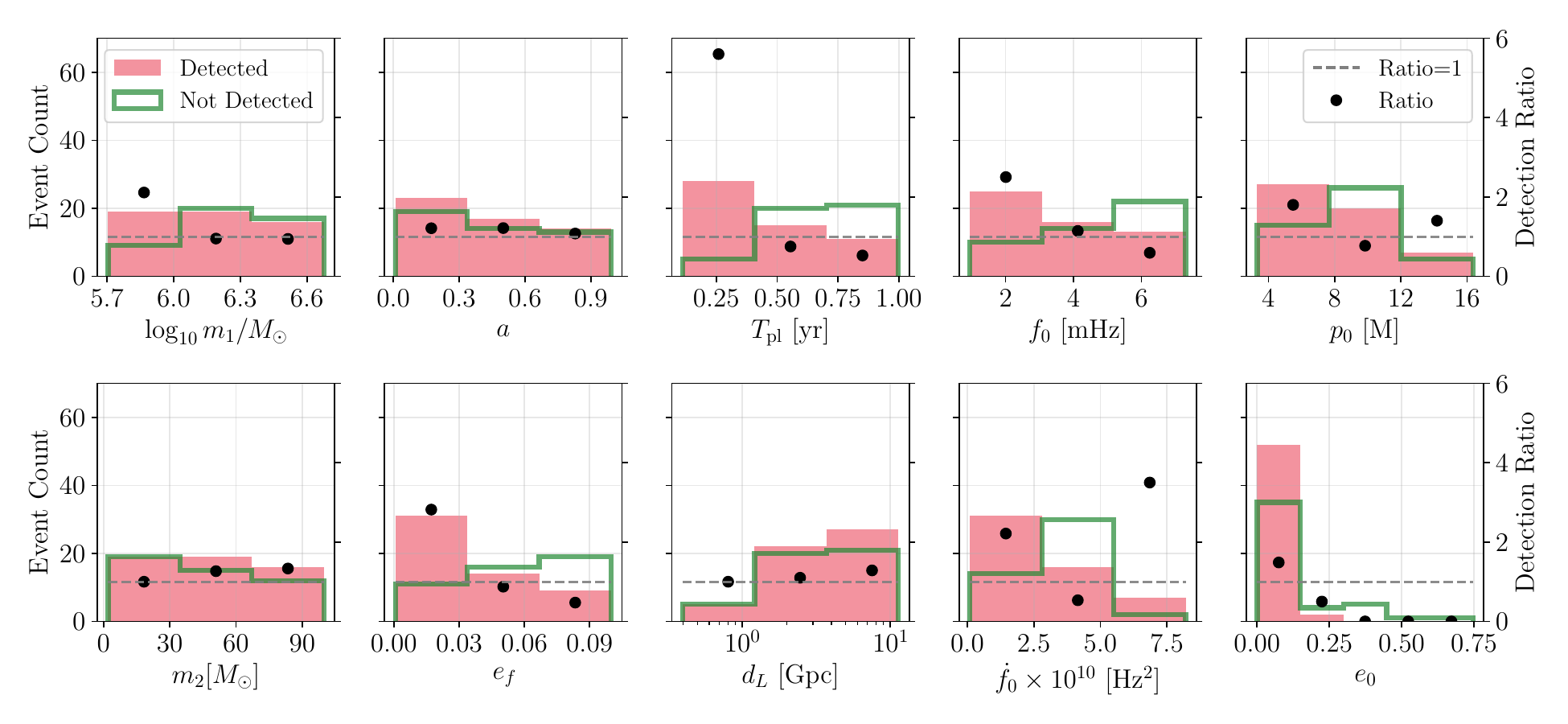}
  \caption{Parameter-space dependence of detection performance for 100 EMRI injections at SNR = 25 with false alarm rate $10^{-4}$. 
  Each panel shows stacked histograms
  where pink filled histograms represent successfully detected signals and the green empty histograms represent missed detections. 
  The detection ratio is shown as black dots and obtained by dividing the counts of the detected and not detected histograms. 
  From top to bottom and from left to right, we show distributions of (log) detector-frame primary mass $m_1$, secondary mass $m_2$, primary dimensionless spin parameter $a$, final eccentricity $e_f$, plunge time $T_{\rm pl}$,  luminosity distance $d_L$, initial frequency $f_0$, initial frequency derivatives $\dot f_0$, initial semi-latus rectum $p_0$, and initial eccentricity $e_0$. The total detection probability for these sources at SNR = 25 with false alarm rate $10^{-4}$ is $\approx 50\%$ as shown in Fig.~\ref{fig:det_prob}.
  }
  \label{fig:parameter_sensitivity}
\end{figure*}

We characterized the detection probability as a function of SNR. However, for fixed SNR or equivalently signal power in the data, different EMRI parameters give rise to different spread of the power in time and frequency and across multiple harmonics. Therefore, it is important to explore how the detection probability varies across the parameter space for fixed SNR. This will also indicate in which regions the search strategy is more or less effective and which regions require different strategies than the proposed one.

For this reason we show in Fig.~\ref{fig:parameter_sensitivity} the detected (pink histograms) and not detected (green) EMRI parameters for SNR=25 at false alarm rate $10^{-4}$. We make this choice because from Fig.~\ref{fig:det_prob} we can see that the detection probability is approximately 50\% giving a reasonable sample size for analysis of detected and not detected systems. To interpret whether the pipeline degrades or improves across each parameter we investigate the ratio of the heights of the detected to not detected histograms heights. If the ratio is approximately constant, it means that the detection probability is approximately constant across that parameter.

Performance is relatively uniform across the mass priors $\log_{10}m_1$ and $m_2$, with a mild degradation at high primary masses and small secondary masses. The detection probability is approximately constant for across spins, whereas it degrades with higher final eccentricities, and higher initial eccentricities. These trends reflect the limitations of our single-harmonic linear chirp approximation.
The degradation at higher plunge times ($T_{\rm pl} \gtrsim 0.75$) is likely due to the increased number of SFTs contributing to the statistic, which raises the detection threshold for a fixed SNR.
The distribution of the initial frequency \(f_0 = f_{\alpha=0}\) indicates that systems with higher starting frequencies are more difficult to detect. These systems evolve more rapidly and may not be well captured by our linear chirp model. The behavior of the frequency derivative can be attributed to low-statistics of 100 realizations. This is supported by checking the behavior across plots at other SNR levels.
We find a relatively constant detection ratio of the luminosity distance $d_L$, whereas the initial semi-latus rectum $p_0$ also shows an alternating pattern.

\subsection{EMRI parameter identification}\label{subsec:parameter_identification}
To refine the identification of EMRI parameters from the recovered frequency track, we perform a dedicated follow-up analysis that combines simulation-based inference and direct evaluation of the detection statistic. We show this using a representative EMRI system 
with intrinsic parameters $m_1 = 1.3379 \times 10^6\,M_\odot$, $m_2 = 27.091\,M_\odot$, $a = 0.8636$, $T_{\rm pl} = 0.902\ \mathrm{yr}$, and $e_f = 0.007086$ at SNR~$=30$, for which the recovered track is shown in Fig.~\ref{fig:frequency_recovery}.
The approach presented in the following is based on empirical validation of convergence and robustness. A systematic evaluation of the follow-up performance across the full parameter space will be presented in future work. An example of Bayesian follow-up framework was explored in \cite{Ashton:2018ure,Tenorio:2021njf,Mirasola:2024lcq,Martins:2025jnq} in the context of continuous gravitational wave analyses.

\begin{figure}
  \centering
  \includegraphics[width=\columnwidth]{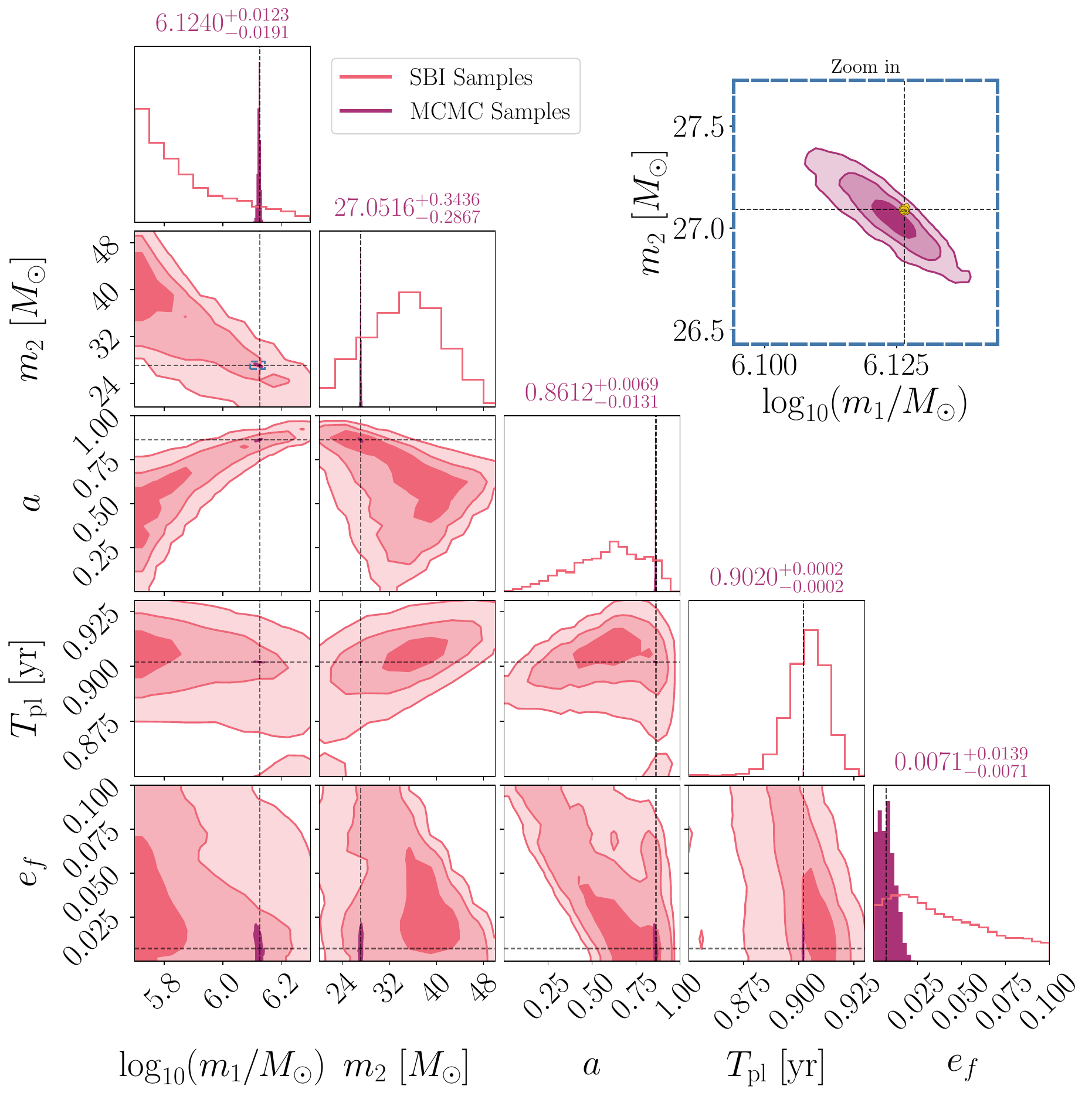}
  \caption{
  EMRI parameter identification for a system with primary mass $m_1 = 1.3379 \times 10^6\,M_\odot$, secondary mass $m_2 = 27.091\,M_\odot$, primary spin $a = 0.8636$, time to plunge $T_{\rm pl} = 0.902\ \mathrm{yr}$, and final eccentricity $e_f = 0.007086$, and SNR = 30. The pink contours show the proposal distribution obtained via SBI using the recovered frequency track from the search pipeline Fig.~\ref{fig:frequency_recovery}. This distribution is used to inform a follow-up analysis.
From $10^6$ samples drawn from the SBI proposal, the highest detection statistic values shown in Fig.~\ref{fig:followup} are used to initialize a follow-up MCMC for refining the EMRI parameter identification. The distribution from the follow-up MCMC is indicated by the violet contours, and an insert for the EMRI component masses is shown in the upper right corner of the figure with 3$\sigma$ credible intervals. 
Labels above the marginal distributions indicate medians and 3$\sigma$ credible intervals of the MCMC samples.
Black dashed lines denote the true injected parameter values.
}
  \label{fig:corner}
\end{figure}

The first stage uses the SBI framework to generate a proposal distribution for the EMRI parameters, conditioned on the {recovered} frequency evolution obtained from the search pipeline. Specifically, we employ a neural density estimator trained on simulated frequency tracks, as described in Sec.~\ref{subsec:sbi}. The resulting proposal distribution, $p(\theta | f^{\rm rec}_\alpha)$, identifies the regions of parameter space most consistent with the observed frequency track at the 1\% relative frequency error level. After ten rounds of inference (approximately ten minutes on a laptop), we obtain the pink distribution shown in Fig.~\ref{fig:corner}. Increasing the number of rounds beyond ten does not significantly change the proposal distribution.
The constraints on the EMRI secondary mass and time to plunge derived from the SBI proposal correspond to relative uncertainties of $5$--$30\%$, with broader constraints on the primary mass, spin, and final eccentricity.
Note that this SBI reconstruction does not directly target the posterior distribution $p(\theta | {\rm data})$, but rather an intermediate distribution derived from the reconstructed harmonic track, $p(\theta | f^{\rm rec}_\alpha)$. Although this distribution in Fig.~\ref{fig:corner} does not include the full data likelihood, it serves as a practical and efficient starting point for further refinement of the EMRI parameters.

\begin{figure}
  \centering
  \includegraphics[width=\columnwidth]{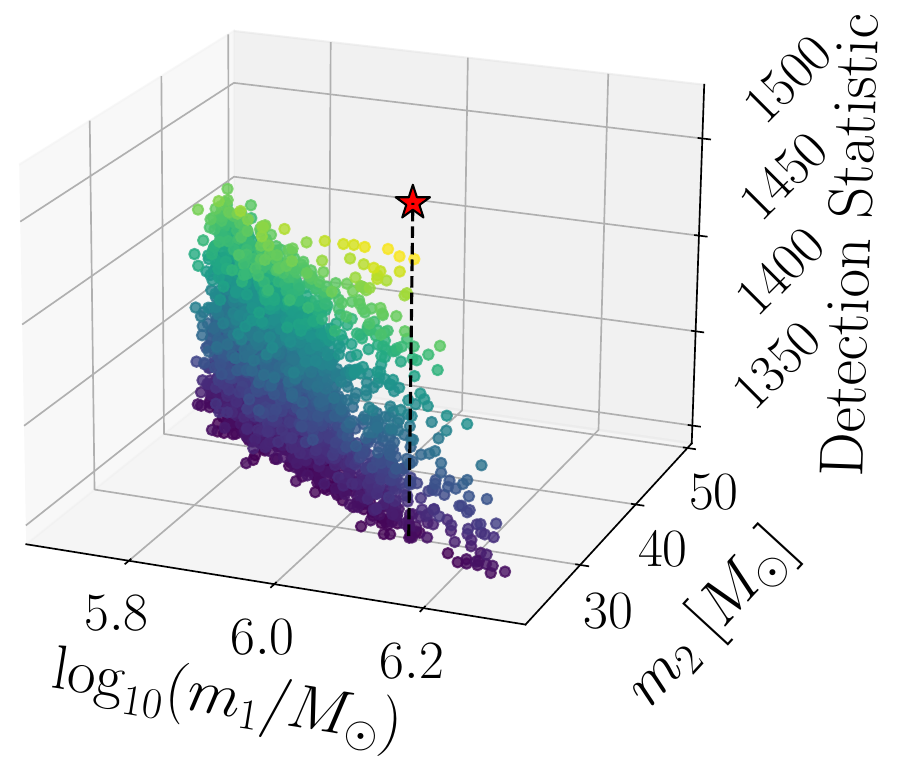}
  \caption{
Detection statistic $\tilde{\mathcal{S}}$ as a function of the primary mass $m_1$ and secondary mass $m_2$, obtained by drawing samples for all EMRI parameters from the SBI proposal shown in Fig.~\ref{fig:corner}. The samples displayed correspond to the top 5000 detection-statistic values out of the $10^6$ evaluated. The maximum detection statistic among the SBI samples yields relative differences of 0.1\% and 1.5\% from the true values for $\log_{10}m_1$ and $m_2$, respectively. The red star marks the best-fit point obtained from the follow-up MCMC sampling, which differs from the SBI maximum by $\tilde{\mathcal{S}} = 25$ and has relative differences of 0.000084\% and 0.3\% from the true values for $\log_{10}m_1$ and $m_2$, respectively.
}
  \label{fig:followup}
\end{figure}

We tested whether stochastic optimization methods could recover the global maximum of the detection statistic by drawing $10^3$ samples from the SBI proposal. However, we found that the algorithm frequently converged to local maxima. To overcome this, we directly mapped the detection-statistic surface over the SBI proposal. Specifically, we drew $10^6$ samples of EMRI parameters $\theta$, and for each sample computed the detection statistic $\mathcal{S}(f_\alpha, \dot{f}_\alpha)$ by generating the corresponding frequency evolution via the EMRI trajectory model, $\theta \xrightarrow{\rm Trajectory} (f_\alpha, \dot{f}_\alpha)$. This procedure incurred a computational cost of approximately 10 milliseconds per sample, resulting in a total runtime of about three hours on a modern laptop using eight parallel processes. The largest detection-statistic values are shown in Fig.~\ref{fig:followup} as a scatter plot over the EMRI component masses. The red star marks the best-fit point from the follow-up analysis discussed below. We note that the sample density around the true parameters is relatively sparse, likely due to the SBI proposal favoring lower primary masses and higher secondary masses. This trend is also visible in Fig.~\ref{fig:corner}, where the SBI density decreases in the direction of the true parameters. Such bias in the proposal distribution may explain why stochastic optimization with fewer samples ($10^3$) failed to converge to the global maximum of the detection-statistic surface.
We found that the highest detection-statistic values were concentrated in a narrow slice of the parameter space (Fig.~\ref{fig:followup}), with a gentle slope toward lower primary masses and a steeper decline toward higher primary masses.
The maximum sample has a relative difference of 1.5\% for $\log_{10}m_1$ and of order 0.1-0.03\% for $m_2, a, e_f, T_{\rm pl}$.
The SBI proposal seems to have guided us along the right correlations of EMRI parameters, but it remains to be seen whether the number of samples used is sufficient to establish a robust global maximum of the detection statistic across different injections.

To further refine the parameter estimates, we select the largest detection statistic samples and use them to initialize a differential evolution optimization as implemented in \texttt{scipy} with options \texttt{strategy=`best1bin', mutation=(0.5, 1), recombination=0.7}~\cite{2020SciPy-NMeth}. The algorithm runs for 100 iterations on 32 walkers and finds candidate solutions that deviate true parameters at sub-percent level.
This process allows to locate the region of maximum detection statistic for starting the final MCMC sampling and avoiding a long burnin phase. The difference between the maximum detection statistic obtained from the 
SBI proposal and the MCMC samples is approximately $\Delta \tilde{S} =25$.

The final stage employs the \texttt{eryn} MCMC sampler~\cite{Karnesis:2023ras} initialized with the differential evolution solutions. 
The MCMC explores the distribution of the EMRI parameters constructed using 
the semicoherent detection statistic [Eq.~\ref{eq:det_stat}] as the likelihood and the priors of Table~\ref{tab:emri_priors}.
We use four temperatures and eight walkers and 1000 iterations to obtain the samples in Fig.~\ref{fig:corner} shown as violet contours.
The {detector-frame} component masses are measured at sub-percent precision at 1$\sigma$ level, with $\log_{10}m_1 = 6.124^{+0.004}_{-0.005}$ (relative precision: $0.070\%$) with $m_2 = 27.1^{+0.11}_{-0.09} M_\odot$ (relative precision: $0.38\%$). 
The primary spin is constrained to $a = 0.861^{+0.002}_{-0.003}$ (relative precision: $0.30\%$). The time to plunge is measured at the $0.008\%$ level, $T_{\rm pl} = 0.902^{+0.0001}_{-0.0001}$ yr. The final eccentricity remains less well constrained, with $e_f = 0.0071^{+0.0052}_{-0.0047}$ (relative precision: $69\%$). These results show that the follow-up analysis refines the EMRI parameter estimates beyond the initial SBI proposal, achieving measurement precisions at the order of sub-percent levels even using a single harmonic track and an approximate detection statistic.

\section{Conclusions}\label{sec:conclusions}

\subsection{Main results}

We developed and characterized a semi-coherent, harmonic-finder pipeline for detecting EMRIs in LISA data. This work addresses the computational challenges associated with EMRI searches by leveraging time-frequency domain techniques, phenomenological frequency evolution modeling through SVD decomposition, and efficient \texttt{jax}-based optimization. We also show how this information can be used to identify the EMRI parameters through a combination of simulation-based inference and direct evaluation of the detection statistic.
Under the assumption of a single eccentric equatorial EMRI in stationary Gaussian noise, our work shows robust detection capabilities across the targeted parameter space. The key findings are:

\begin{itemize}

\item \textbf{Pipeline performance.} Our method achieves 80\%  (99\%)
detection probability at SNR $\approx$ 30 
for a false alarm probability of $10^{-4}$ ($0.5$). 
For the same SNR, we can determine the frequency evolution with 
relative precision of $|\delta f / f| \sim 10^{-3}$.

\item \textbf{EMRI identification.} We showed how to construct proposals based on frequency tracks. This level of frequency precision of 1\% allows the simulation based inference to create a proposal with the secondary mass and time to plunge with 30\% and 1\% precision, whereas the other parameters remain unconstrained. It is only through a frequency evolution based on the real EMRI trajectory that we constrain the parameters to sub-percent precision.
Finally, we showed that the detection statistic of Eq.~(\ref{eq:det_stat}) can be used to refine the EMRI parameters to sub-percent precision on component masses, primary spin, and time to plunge.

\item \textbf{Parameter space coverage.} The detection probability remains constant across the parameters $\log_{10}(m_1)$ and $m_2$, and degrades with increasing eccentricities $e_0, e_f$, reflecting limitations of our single-harmonic, linear-chirp approximation. A decrease in sensitivity is also observed for longer plunge times $T_{\rm pl}$ due to signal power spreading over time at fixed SNR.

\item \textbf{Computational efficiency.} Our pipeline processes one year of data in a wall-clock time of about 1 hr {on a GPU}, requiring $\approx 10^7$ detection statistic evaluations. Our follow-up parameter identification procedure refines EMRI parameters within hours on a CPU.

\end{itemize}

\subsection{Future improvements}

Several enhancements could improve both performance and applicability of our method.

\begin{itemize}
\item \textbf{Multi-harmonic extension.} Including multiple harmonics would improve sensitivity for eccentric systems and provide additional parameter constraints. The time-frequency framework naturally accommodates multi-mode searches with proportional computational-cost scaling. Possible approaches to include higher harmonics could follow Refs.~\cite{Wadekar:2023kym,Roy:2019phx}.

\item \textbf{SBI harmonic proposal.} In this work, we trained the SBI proposal conditioned on an observation. Future work could train the SBI directly on the full EMRI parameter space, allowing an amortized proposal. This would reduce the computational cost of the follow-up analysis in the case of multiple detections. 
Furthermore, it remains to be seen whether it is possible to train a network to predict the non-local secondary maxima in the likelihood surface of Ref.~\cite{Chua:2021aah} where the relative difference in frequency is of order of $10^{-6}$.

\item \textbf{EMRI identification across the parameter space.} In this work, we showed EMRI parameter identification for a single representative system. Future work should systematically evaluate the follow-up performance across the full parameter space, including varying SNR levels and eccentricities. The goal would be to show that the detection statistic introduced in this work is sufficient to obtain EMRI parameters at the sub-percent level as done in the considered case.

\item
\textbf{Higher-order waveform approximations.} Incorporating quadratic chirp terms and amplitude evolution could extend validity closer to plunge and improve accuracy for rapidly evolving systems. In particular, relaxing the linear chirp approximation would allow for longer SFT durations and improved frequency reconstruction.
\end{itemize}

\begin{figure*}
  \centering
  \includegraphics[width=1.98\columnwidth]{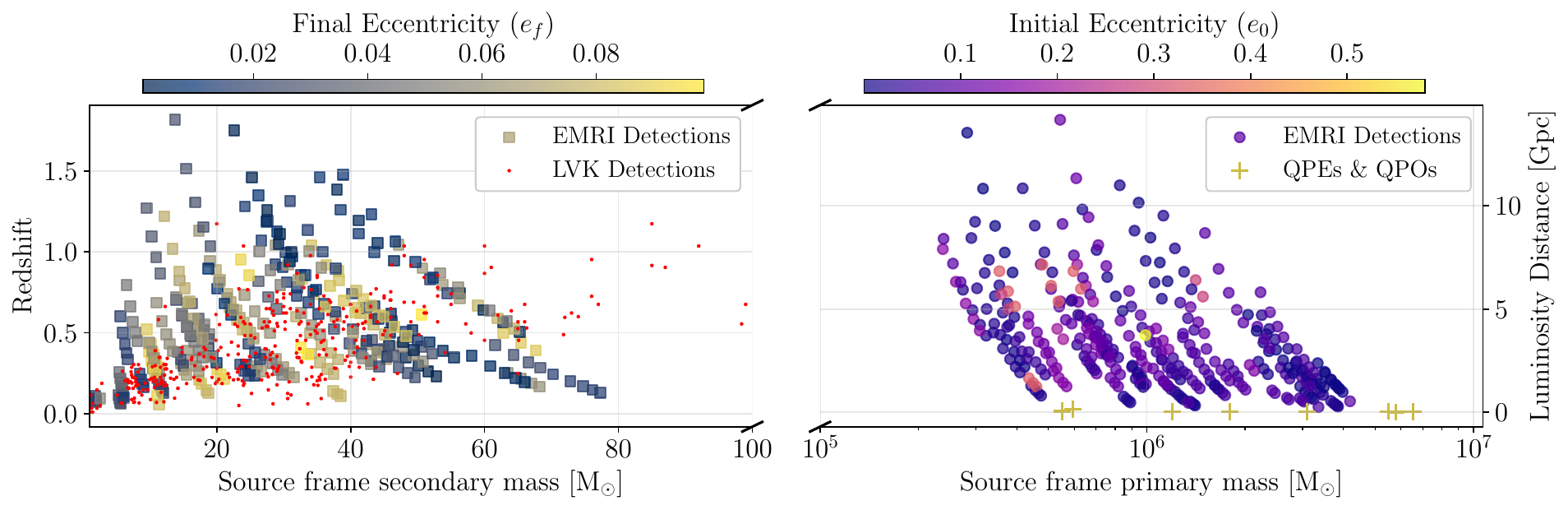}
  \caption{Parameter-space dependence of the EMRI systems detected in this work with false-alarm rate $10^{-4}$ and $\mathrm{SNR} = [20, 25, 30, 35, 40]$. 
  The $y$-axes show redshift (left) and luminosity distance in Gpc (right). 
  The $x$-axes show the EMRI source-frame secondary mass $m_2/(1+z)$ (left, squares) and source-frame primary mass $m_1/(1+z)$ (right, circles). The color scales indicate the final eccentricity $e_f$ (left) and the initial eccentricity $e_0$ (right).
  For reference, we show as red dots in the left panel the masses (both primary and secondary) and redshift of the current LIGO/Virgo/KAGRA detections \citep{LIGOScientific:2018mvr,LIGOScientific:2020ibl,KAGRA:2021vkt,
LIGOScientific:2021usb,LIGOScientific:2025slb,LIGOScientific:2019lzm,KAGRA:2023pio,LIGOScientific:2025snk}. 
  We also show the QPE and QPO samples collected in Ref.~\cite{Kejriwal:2024bna}.
  }
  \label{fig:horizon_distance}
\end{figure*}

\subsection{Astrophysical implications}

We qualitatively discuss the astrophysical implications of our results. While a complete assessment must ultimately include overlapping sources and realistic instrumental noise, we focus on the representative subset of the EMRI parameter space (Table~\ref{tab:emri_priors}) detected for sources with different  signal-to-noise ratios obtained from a single data stream channel.

Figure~\ref{fig:horizon_distance} shows the distribution of detectable EMRI systems at a false-alarm probability of $p_{\rm FA}=10^{-4}$ and $\mathrm{SNR} = [20, 25, 30, 35, 40]$. The left panel displays the source-frame secondary masses as a function of redshift (left y-axis) and luminosity distance (right y-axis), color-coded by the final eccentricity $e_f$. The right panel presents the source-frame primary masses, color-coded by the initial eccentricity $e_0$. The observed skewness in the mass distributions arises from the $(1+z)$ correction used to convert detector-frame to source-frame masses. 

For reference, we show the primary and secondary masses (median) and redshifts of current LIGO/Virgo/KAGRA (LVK) detections~\citep{LIGOScientific:2018mvr,LIGOScientific:2020ibl,KAGRA:2021vkt,
LIGOScientific:2021usb,LIGOScientific:2025slb,LIGOScientific:2019lzm,KAGRA:2023pio,LIGOScientific:2025snk} as red dots in the left panel. 
EMRI detectable systems extend to higher redshifts than LVK sources for secondary masses smaller than $20\,M_\odot$.
Although EMRIs and LVK sources originate from distinct formation channels, the secondary object in an EMRI may still form through stellar evolution before capture, providing a complementary perspective on stellar-mass compact object populations in galactic centers.
In particular, the secondary-mass distribution reveals detectability of lighter remnants ($m_2 \lesssim 30\,M_\odot$) out to $z \sim 1.5$, probing compact object demographics in galactic nuclei near cosmic noon—the epoch of peak star formation and galaxy merger activity around $z \sim 1$--2~\cite{2020ARA&A..58..661F}.
We also notice that our prior range is limited to source-frame secondary masses of $80\,M_\odot$, and future analyses should be extending this range.

EMRIs formed through disk-assisted channels~\cite{Pan:2021ksp}, sometimes called ``wet EMRIs,'' are expected to occur in a fraction $\sim 10^{-2}$--$10^{-1}$ of galactic nuclei. In these systems, interactions between the compact object and the surrounding accretion disk can lead to high-energy electromagnetic emission, making them promising candidates for multi-messenger observations. 
Two interesting classes of such events are the so-called quasi-periodic eruptions (QPEs) and quasi-periodic oscillations (QPOs).
The former are semi-regular bursts observed in soft X-ray bands and the latter are less abrupt events due to oscillations in the X-ray flux of active galactic nuclei (AGNs). 
Under the assumption that these phenomena are powered by EMRIs, we show
in the right panel of Fig.~\ref{fig:horizon_distance} the massive black hole masses and redshifts of QPEs and QPOs~\cite{Miniutti:2019fqr,Arcodia:2024krd,Giustini:2020gex,Gierlinski:2008yz,Arcodia:2021tck,Lin:2013wqa,Pasham:2018bkt,2024SciA...10J8898P,2024NatAs...8..347G} from Table~I of~\cite{Kejriwal:2024bna}. 
The EMRI sources considered in this work are detected at higher redshifts than these QPE and QPO events~\cite{Kejriwal:2024bna}. 
Future work will need to investigate the sky localization precision of follow-up EMRI gravitational-wave analyses required to enable electromagnetic observations for multi-messenger studies of these systems.
We notice that our prior does not include source-frame primary masses larger than $5\times 10^6 \, M_\odot$ so future works will need to assess the detectability of EMRIs with large primary masses $\sim 10^7M_\odot$, where the impact of overlapping galactic binaries might play a crucial role.

Because our pipeline is optimized for low-eccentricity systems, quantifying detection prospects across realistic astrophysical populations remains an open question. Wet EMRIs—formed via migration through AGN accretion disks~\cite{Lyu:2024gnk,Pan:2021ksp}—are expected to have quasi-circular orbits ($e_f \lesssim 0.01$) around rapidly spinning primaries ($a \gtrsim 0.9$), as gas drag circularizes the inspiral and aligns its inclination. Our method performs well in this low-eccentricity regime, however the impact of inclination on the search strategy has not yet been considered. This will need to be evaluated once fully relativistic generic waveforms become available.

In contrast, EMRIs formed via the two-body relaxation channel in nuclear star clusters are expected to be highly eccentric and will likely require a dedicated search strategy~\cite{Mancieri:2025cmx,Amaro-Seoane:2007osp}. Assessing the minimum SNR required for a confident detection based on a single harmonic would be an important next step toward characterizing this pipeline's performance in the high eccentricity regime.

\subsection{Implications for the LISA global fit}
Modern global-fit architectures~\cite{Littenberg:2020bxy,Speri:2022kaq,Littenberg:2023xpl,Katz:2024oqg,Strub:2024kbe,Deng:2025wgk} rely on well-designed proposal distributions to guide both the search and parameter estimation for various sources. For example, previous studies have employed proposal distributions based on normalizing flows to enhance the convergence of galactic binary sampling~\cite{Korsakova:2024sut}.

In the context of the LISA global fit, the pipeline introduced in this work can identify EMRI candidates and, combined with the SBI framework, provide constrained parameter space regions for follow-up analyses using full-waveform templates.
Our analysis showed an example of such a follow-up using the statistic defined in Eq.~(\ref{eq:det_stat}). 
{The search and identification pipelines developed here do not rely on waveform templates; instead, they exploit only the relationship between EMRI parameters and their frequency evolution}.  %
This approach significantly reduces computational cost and enables scalable candidate identification across a broad parameter space.
Further investigation is required to assess whether the statistic defined in Eq.~(\ref{eq:det_stat}) is sufficient for robust EMRI identification throughout the full parameter space. 

Moreover, the long-duration nature of EMRI signals makes them more likely to be affected by non-stationary noise and instrumental artifacts~\cite{2024PhRvD.110d2004A,Castelli:2024sdb}. Recently, Ref.~\cite{Bandopadhyay:2025fyx} showed that for stellar origin black hole binaries the detection statistic of Eq.~(\ref{eq:det_stat}) is robust against non-stationary noise. Understanding how these effects impact EMRI detection and parameter estimation will be crucial.

Galactic binaries and massive black hole binaries will influence EMRI detection differently based on their distinct characteristics in the time-frequency plane~\cite{Khukhlaev:2025xiz}. Future work should investigate whether the detected frequency tracks could be used to mask signals and enable targeted EMRI follow-up. Characterizing and leveraging these multi-source interactions will be crucial for integrating a robust EMRI search capability into the LISA global fit.

\begin{acknowledgments}
We thank Jonathan Gair, Elena Maria Rossi and Philippa Cole %
for discussions.
L. S. thanks Gijs Nelemans for providing the computational resources that enabled the large-scale simulations and data analysis presented in this work. 
L. S. is supported by the  European Space Agency Research Fellowship programme.
R.T. and D.G. are supported by 
ERC Starting Grant No. 945155–GWmining, 
Cariplo Foundation Grant No.~2021-0555,
MUR PRIN Grant No.~2022-Z9X4XS, 
Italian-French University (UIF/UFI) Grant No.~2025-C3-386, 
MUR Grant “Progetto Dipartimenti di Eccellenza 2023-2027” (BiCoQ), 
and the ICSC National Research Centre funded by NextGenerationEU. 
C.E.A.C-B. is supported by UKSA Space Agency grant UKRI971.
D.G. is supported by 
MUR Young Researchers Grant No.~SOE2024-0000125,
and MSCA Fellowship No.~101149270–ProtoBH.
This work used the Dutch national e-infrastructure with the support of 
the SURF Cooperative using grant no. EINF-10027.
Computational work was performed 
at CINECA with allocations through INFN and the University of Milano-Bicocca, 
and at NVIDIA with allocations through the Academic Grant program.
The authors thank participants and organizers of the  \textit{Scientific Machine Learning for Gravitational Wave Astronomy} workshop held at ICERM (Providence, RI), which is supported by NSF Grant No. DMS-1929284.
This research has made use of data or software obtained from the Gravitational Wave Open Science Center (gwosc.org), a service of the LIGO Scientific Collaboration, the Virgo Collaboration, and KAGRA.
This work makes use of the Black Hole Perturbation Toolkit, \texttt{few}  \cite{Chua:2020stf,Katz:2021yft,Speri:2023jte,Chapman-Bird:2025xtd}, \texttt{numpy}~\cite{harris2020array}, \texttt{matplotlib}~\cite{Hunter:2007}, \texttt{scipy}~\cite{2020SciPy-NMeth}, \texttt{jax}~\cite{jax2018github}, \texttt{optax}~\cite{deepmind2020jax}, \texttt{interpax}~\cite{conlin_2025_17378822}, and \texttt{pastamarkers}~\cite{2025arXiv250323126P,2024arXiv240320314P}.
\end{acknowledgments}

\bibliography{emri_search}

\end{document}